\documentclass[11pt]{article}
\newtheorem{hb}{Holevo Bound}
\begin{document}
\begin{titlepage}
\vspace{4cm}
\begin{center}{\Large \bf Achieving Holevo Bound in Quantum Measurement}\\
\vspace{1cm}Nima Lashkari\footnote{Corresponding Author\\email:nlashkari@gmail.com}\\
\vspace{1cm}Department of Physics, Sharif University of Technology, 14588-89694, P.O.Box: 11365-9161,\\ Tehran, Iran
\end{center}
\vskip 3cm
\begin{abstract}
We show that information gain in a qubit measurement is optimal under a Von Neumann measurement.
For an initially mixed apparatus kept in touch with a qubit, the conditions for achieving the
equality sign of Holevo bound on the information accessible to apparatus are derived.
These constraints can be identified as the conditions for the optimization of information gain in a
 qubit measurement. At the end, we will generalize the idea to qudit measurements using a phase-shift gate.
\end{abstract}
\end{titlepage}
\vskip 3cm
\section{Introduction}
\label{sec:1}
A fundamental statistical interpretation lies in the heart of quantum mechanics. Given an unknown state,
 there is no way to figure out what the state is because of the so-called collapse postulate.
 The problem of state determination remains a challenge even in statistical ensembles \cite{hol}.
  Having prepared an infinite ensemble in the state $\psi$, we can learn about the unknown state
  from the frequency of measurement outcomes. Information is transformed from the system to an
  apparatus via interaction\cite{wheel}. This interaction entangles apparatus with the system
  and extracts the quantum information initially contained in quantum system. In the next stage
   of measurement procedure, the apparatus spreads out this information in its surrounding environment.
   It is eventually because of the numerous degrees of freedom of environment states, that we consider
   this information transfer is irreversible. Usually information transfer from the system to apparatus
   is not perfect and some information is lost on the way. In this paper, we intend to optimize the
   information transfer rate.
In an ideal measurement, all that is accessible to us are the probabilities $|c_i^2|$ in the extension
$\psi=\sum_i c_i|\psi_i\rangle$.Let's denote probabilities $|c_i^2|$ with the random variable $X$.
In an imperfect measurement when some information is lost during the transfer, we read the random variable
 at the apparatus. In this case, the mutual information $H(x:y)$ is a criterion for the amount of transferred
 information. A closer $H(x:y)$ to $H(x)$ is a sign of less information loss, and a more successful measurement.
 Therefore, we should plan measurements which maximize this quantity. One way is to force the system and the apparatus to undergo some appropriate interactions. Here we are going to find the general form of the interactions that optimize information transfer in qubit measurements. It is of special interest because of its vast applications in quantum computation, information and quantum communication. Just as an illustration, in any quantum circuit any line must end to a qubit measurement and it is obvious that the success of a quantum circuit highly depends on a successful final measurement. We are going to find interactions which transfer information optimally by achieving the so-called Holevo bound. At the first step it is done for two level systems -qubits- and then generalize to any d-level system.

\section{Model}
\label{sec:2}
The model consists of a system and two ancillas, an apparatus and environment. The apparatus is kept in touch with environment and, in general, it has a mixed density matrix. Our goal is to construct the closest random variable to $|c_i|^2$ or in another word, optimization of information gain in a quantum measurement. The random variable  is read with the operation of appropriate POVM elements on the apparatus state and discriminating between them. The situation is quite analogous to the classical communication over quantum channels.  Interaction encodes variable $X$ on the apparatus density matrices with probability distribution $|c_i^2|$ just like the case of a quantum channel where Alice encodes information content of random variable $X$ in density matrices $\rho_i$ with the same probability distribution of $x_i$. And at the end of channel Bob reads $Y$ with POVM measurements on $\rho_i$. In our model Alice is the system and Bob is the observer, while the encoding procedure is the entanglement of the system and apparatus
\begin{hb}
Consider a prepared state $\rho_x$ while $x=0,...,N$ with probabilities $p_1,p_2,p_N$. The Holevo bound states that for any POVM measurement done on these matrices the accessible information is bounded by
\begin{equation}
    H(x:y)\leq S\Big(\sum_x\rho_x p_x\Big)-\sum_xp_xS(\rho_x)
\end{equation}
The equality holds if and only if all $\rho_x$ commute.\cite{hol2, fuch, cerf, schum}.
\end{hb}

As can be seen from the above bound, in our problem determining the interactions which bring the initial density matrix of the apparatus to commuting matrices $\rho_i$ with probabilities $|c_i|^2$ optimizes the information transfer.

Consider a system in the state $\psi=\sum_i c_i|\psi_i\rangle$ and an apparatus in a totally mixed state $\rho=\sum_k r_k|r_k\rangle\langle r_k|$. As the apparatus is brought into contact with the system, they interact through a unitary evolution and consequently entangle
\begin{equation}
    \rho\oplus|\psi\rangle\langle\psi|\to\sum_{i,j} c_i c_j^* r_k U|i\rangle|r_k\rangle\langle j|\langle r_k|U^\dagger
\end{equation}
Suppose that the interaction has a form like \cite{ved}
\begin{equation}
    U|i\rangle|r_k\rangle=|i\rangle|r^i_k\rangle
\end{equation}
Since $U$ is unitary, we have the orthogonality condition
\begin{equation}
    \forall\:i,k,k^\prime: \langle r_k^i|r_{k^\prime}^i\rangle=\delta_{k,k^{\prime}}
\end{equation}
Taking trace over the system, the density matrix of apparatus leaves
\begin{equation}
    \rho^\prime=\mathrm{tr_s}\Big(\sum_{i,j}\sum_k c_i c_j^* r_k|i\rangle|r_k^i\rangle\langle j|\langle r_k^j|\Big)=\sum_i |c_i|^2\Big(\sum_k r_k|r_k\rangle\langle r_k|\Big)
\end{equation}
Call the term in the parenthesis $\rho_i$. We have
\begin{equation}
    \rho^\prime=\sum_i |c_i|^2\rho_i
\end{equation}
We are now in the position where the information content of random variable $x$ is encoded in $\rho^\prime$ and it is the observer's turn to measure and discriminate between $\rho_i$. As a result, the problem of optimization of information gain reduces to the problem of finding transformations $U$ which can reach Holevo bound for any $|\psi\rangle$ and $\rho$.

From the Holevo bound we know that in order to reach the bound for the accessible information all $\rho_i$ must commute. In the following sections, we will discuss conditions on $U$ to reach Holevo bound.
\section{Qubit Measurement}
\label{sec:3}
We will see that, under Von Neumann condition for measurement interaction \cite{von} the bound is always achieved for qubit measurements.
\begin{equation}
    \forall\:i,j,k: \langle r_k^i|r_k^j\rangle=\delta_{ij}
\end{equation}
Suppose $\big\{|r_a^1\rangle,|r_b^1\rangle\big\}$ and $\big\{|r_a^2\rangle,|r_b^2\rangle\big\}$ are two orthonormal bases for the two dimensional Hilbert space $H$. Under the Von Neumann conditions $\langle r_a^1|r_a^2\rangle=0$ and $\langle r_b^1|r_b^2\rangle=0$ the bases are the same up to some phase factors. The proof is rather trivial. Since $|r_a^1\rangle$ is orthonormal to both $|r_a^2\rangle$ and $|r_b^1\rangle$ they just can differ in a phase. This is also true for $|r_a^1\rangle$ and $|r_b^2\rangle$. Therefore, these bases are the same up to some phases.

Accordingly, under the condition of Von Neumann interaction matrices are diagonalizable in the same basis. As a result, they commute.
In order to maximize the information transfer, the system and the apparatus must undergo a Von Neumann interaction in a qubit measurement.
\section{Generalization to D-level Systems}
\label{sec:4}
In this section, we shall find a unitary evolution  that can extract maximum information from a d-level system. First, let's start with a three level system. In analogy to the lemma proved in the last section, it can be proved with a similar proof that Von Neumann interactions are the appropriate interactions in the case of a three level system, too. While for higher level system this is not always true. It also turns out to be a difficult task to find the general form of the evolutions which lead to commuting $\rho_i$ in bigger Hilbert spaces. Instead, we will prove shift-gate to be one of the desired unitary transformations for arbitrary level systems.
    The main idea originates form the CNOT gate -controlled NOT-, a famous transformation in quantum computation and information.  CNOT operates on two qubits, one control and one target. The transformation is such that if the control qubit is $|1\rangle$ the target is flipped while in the other case neither the control nor the target qubit change. This gate is one of the Von Neumann interactions for the qudits that  transfer information optimally. This is easy to check. In a CNOT gate with the system state as its control qubit and the apparatus state as its target qubit, the final state of target flips as the control flips which is the same as condition $\langle x^0|x^1\rangle=0$. Hence, the information is transferred from the system to apparatus.

The most direct generalization is to make a shift-gate acting like
\begin{equation}
    U|i\rangle|r_k\rangle=|i\rangle|r_k\oplus i\rangle
\end{equation}
Here summation is modulo $d$. The corresponding matrix of this evolution is
\[U= \left( \begin{array}{ccc}
D^0 & \cdots & 0 \\
\vdots & \ddots & \vdots \\
0 & \cdots & D^{d-1}\end{array} \right)\]
where $D^k$ is a $D\times D$ matrix defined according to $D_{ij}^k=\delta_{i,(j+k\:\mathrm{mod}d)}$

As a result of (8), the bases $\big\{|r_k\oplus i\rangle\big\}$ in which matrices $\rho_i$ are diagonal (4) are the same, but a permutation on labels of basis vectors. This means that all $\rho_i$ commute. Accordingly, shift-gate evolution on system and apparatus of arbitrary dimension achieves Holevo bound. This is what we were looking for to optimize information gain.
\section{Discussion}
\label{sec:5}
We have presented an optimization of information gain in a quantum measurement by controlling the interaction of system and apparatus.  We found a group of interactions which are optimal according to the Holevo bound. Open problems to be solved are finding the appropriate POVM elements that act on the final apparatus state.

Acknowledgement: At the end, I would like to thank Salman Abolfath Beigi for useful discussions.

\end{document}